\documentclass[twocolumn,showpacs,showkeys]{revtex4}

\usepackage{epsfig}
\usepackage{amsmath}
\usepackage{bm}% bold math

\begin{document} 

\title{Stabilization of magnetic polarons in antiferromagnetic
semiconductors by extended spin distortions.}

\author{J. Castro}
\author{I. Gonz\'alez}
\email{faivan@usc.es}
\author{D. Baldomir}

\affiliation{Departamento de F\'{\i}sica Aplicada, Universidade        %
de Santiago de Compostela,\\ E-15706 Santiago de Compostela, Spain.}

\begin{abstract}

We study the problem of a magnetic polaron in an antiferromagnetic
semiconductor (ferron). We obtain an analytical solution for the
distortion produced in the antiferromagnetic structure due to 
the presence of a charge carrier bound to an impurity. The region
in which the charge carrier is trapped is of the order of the
lattice constant (small ferron) but the distortion of the magnetic
structure extends over a much larger distance.
It is shown that the presence of this distortion makes the ferron
more stable, and introduces a new length scale in the problem.

\end{abstract}

%Add the option ``showpacs'' into documentclass to show them.

%Add the option ``showkeys'' into documentclass to show them.

\maketitle

\section{Introduction.}

Quite a while ago Nagaev \cite{Nag67} has shown that the minimal       %
energy for a charge carrier moving in an antiferromagnetic background 
is obtained when the electron modifies the magnetic background and 
is self-trapped in a region with canted antiferromagnetic or 
ferromagnetic order. The term  ferron (a magnetic polaron in an 
antiferromagnetic background) was coined there to name this new 
quasi-particle.
Depending on the radius of the self trapping region, it is possible to
differentiate between small ferron, localized in a region of the order
of the lattice constant, and large ferron localized at larger scales.  
Small bound ferrons are expected to be typical of the low doping
region of the phase diagram of manganites, when the material is
antiferromagnetic and insulating. 

There are also experimental data that confirm the existence of bound
ferrons in antiferromagnetic semiconductors such as underdoped
manganites. A review of them can be found in \cite{Nag01R, Dag01}. 
In reference \cite{Hen98}, a liquid-like spatial distribution of
magnetic droplets in La$_{1-x}$Ca$_{x}$Mn$_{3}$ with $x=0.05,\,0.08$ 
is reported. Ferromagnetic rich-hole droplets with a diameter of
$4-5$ lattice units isotropically distributed with a mean distance of
$9$ lattice units among them are observed in an antiferromagnetic
poor-hole background. Also it is reported that these magnetic droplets
are coupled together through the antiferromagnetic background. In
reference \cite{Allo97}, a $^{55}$Mn NMR study on the same compound 
within the doping range $x<0.23$, confirms the electronic phase
separation and report the existence of thin boundaries between
antiferromagnetic and ferromagnetic domains.
These magnetic droplets contain a number of conduction electrons larger
than one (about 30), but this is not essential in describing the spin
distortion that they create in the antiferromagnetic background. 

Almost in all the calculations of the ferrons it  was assumed that the
region of the distortion of the magnetic structure coincides with the
range of localization of the electron. Often these regions were simply
taken as spheres of radius $R$, to be determined self-consistently.
However, as first pointed out by De Gennes \cite{DeG60}, the distortion
of the magnetic order around a magnetic defect (i.e. ferron) may decay
slowly with distance. In this paper we study what would be the ``back 
effect'' of this slowly decaying magnetic distortion on the
conditions of the electron localization and on the properties of the
resulting self-trapped state. First numerical calculations addressing
this problem have been carried out recently by Nagaev  \cite{Nag01b}. 
In this paper the magnetic anisotropy, always present in real materials,
was neglected, and he did not manage to obtain the results in a closed
form and did not get proper estimates for the radius and the energy of
the ferron taking this effect into account. 

In this article, we obtain the analytical solution for the distortion
created by a localized conduction electron trapped in a region of the
order of the lattice constant in an one-dimensional antiferromagnetic
semiconductor. The main virtue of our calculation is than reveals the
existence of a region surrounding the trapping region which acts as an 
antiferromagnetic domain wall, as it was anticipated in reference 
\cite{Nag01b}. We calculated the properties of the ferron in this 
situation and show that the account of an extended magnetic distortion
around it  actually leads to an increase of the ferron stability and
its binding energy. Our result offers a key to understand the features
of the phase diagram of antiferromagnetic semiconductors, such as
manganites at the small doping range $x \alt 0.1$.

\section{Results}

To study the problem of the magnetic polaron formation, we start       %
from the Hamiltonian proposed by Nagaev \cite{Nag01b}. We analyze the
one-dimensional case because it can be treated analytically and its
solution provides a full physical insight into the problem. Estimates
to the three-dimensional case are given below. We consider an one-dimensional
chain of magnetic ions along the $y$-axis, described by the double exchange
(or Vonsovsky s-d) model. An uniaxial magnetic anisotropy term is added
with $x$ being the easy axis. The magnetic structure of the  magnetic 
ions without the conduction electron is represented by two sub-lattices
with the spin up along the easy axis (that means $S^{x}=S$) for ions at
even sites ($g=2n$), and spin down (that means $S^{x}=-S$) for ions at 
odd sites ($g=2n+1$). All distances are measured in terms of the lattice 
constant. A non-magnetic donor impurity is added to the chain at the
point $y=-1/2$. Its conduction electron is bound to it and it can 
jump between their two magnetic neighbors (bound ferron). This disturbs
the pure antiferromagnetic order along the chain, even though the
conduction electron is trapped only on these two magnetic ions. 
The Hamiltonian of the system can be represented by:

\begin{eqnarray}\label{ eq:1}
H_{sd}&=&
-t\left(a^{+}_{-1,\sigma}a_{0,\sigma}+a^{+}_{0,\sigma}a_{-1,\sigma}\right)
\nonumber\\
& &-A \sum\limits_{ g=-1,0}\left(\vec{s} \vec{S_{g}}\right)_{\sigma,\sigma'}
a^{+}_{g,\sigma}a_{g,\sigma'}\nonumber\\
& &-I \sum\limits_{g}\vec{S_{g}}\vec{S}_{g+1}
-K' \sum\limits_{g}\left(S^{x}_{g}\right)^{2}
\end{eqnarray}
where $a^{+}_{g,\sigma},a_{g,\sigma}$ are the conduction electron
operator corresponding to the site $g$ and spin projection $\sigma$,
$\vec{s}$ the conduction electron spin operator, $\vec{S_{g}}$ is the
spin operator of the magnetic ion at site $g$ (d-spins). The d-d 
exchange integral $I$ is assumed negative in order to get the 
antiferromagnetic ordering. The anisotropy constant $K'$ is considered 
positive. The energy of Coulomb interaction between the conduction 
electron and the impurity is an additive constant and, for this reason, 
omitted in the calculation.

Depending on the relative value of the parameters $W=2zt$ and $AS$,
$z$ being the number of first neighbors and $S$ the magnitude of the
d-spin, we have two different situations. In the case $W>>AS$, we talk
about a wide-band semiconductor. In the case $W<<AS$, we talk
about a double exchange semiconductor.

Our goal is to obtain an expression for the magnetic energy of the
system of d-spins, both in the case of wide-band and double exchange
semiconductors.
As $S\ge2$ for the compounds of interest (typically, manganites),
the d-spins are considered classically. Their
orientations are described in a coordinate system centered in the
position of each magnetic ion. As it was stated above, it is assumed that 
the conduction electron is in the lowest bound state in the space spanned 
by the operators of the sites $g=-1,0$. Also we assume that the magnetic 
moment of the ferron is directed along $z$-axis, that is, the 
conduction electron spin acts as an effective magnetic field along the 
$z$-axis for the d-spin system. In this case, the following symmetries 
hold for the d-spin system: $S^{y}_{g}=0$, and 
$S^{z}_{g}=S^{z}_{-\left(g+1\right)}$. Then
$\vec{S}_{g}=S\left((-1)^{g}\sin\theta_{g},0,\cos \theta_{g}\right)$. 
It is important to notice that here, although the d-spin lies in the 
$z-x$ plane, the angle $\theta$ is not the polar angle, but the
spherical coordinate.

To obtain the magnetic energy of the d-spin system the electronic part of 
the Hamiltonian (\ref{ eq:1}) must be averaged out. To do this, we assume
that the electronic wave function is the ground state wave function for
the dominant term in the electronic Hamiltonian, i.e. the hopping term in 
the wide-band case, and the s-d exchange term in the double exchange case. The
other term is treated as a perturbation \footnote{An exact solution for 
the full electronic Hamiltonian is avaliable, but we prefer to use this
approximation, by proposes of clarity. The exact solution can not be used
in the analytical approach we made here.}.
In the case of wide-band semiconductor, the hopping term is diagonalised 
and its ground state is $\left| \Phi \right\rangle=\frac{1}{\sqrt{2}}
\left(a^{+}_{-1,1/2}+a^{+}_{0,1/2}\right)\left|0\right\rangle$.
In the case of double exchange semiconductor, the ground state is the same
but the operators $a^{+}_{g,1/2}$ being the operators with the spin projection
along the direction of the vector $\vec{S}_{g}$, instead of the laboratory
$z$-axis. We treat the case of a wide-band semiconductor first. Then the
magnetic energy of the d-spin system is:

\begin{eqnarray} \label{ eq:2}
E=J \sum\limits_{ g}\cos \left(\theta_{g}+\theta_{g+1}\right)
\nonumber\\
-L[\cos \theta_{-1}+\cos \theta_{0}]
-K \sum\limits_{ g} \sin^{2}\theta_{g}-t
\end{eqnarray}
where:
$J=-IS^{2}$, $L=AS/4$, and $K=K'S^{2}$.

Minimizing the equation (\ref{ eq:2}) with respect to the angles
$\theta_{g}$, a set of non-linear equations is obtained: 

\begin{eqnarray} \label{ eq:3}
J\sin\left(\theta_{g}+\theta_{g+1}\right)+
J\sin\left(\theta_{g-1}+\theta_{g}\right)-\nonumber\\
-L\sin\theta_{g}[\delta_{g,-1}+\delta_{g,0}]
+K\sin\left(2\theta_{g}\right)=0
\end{eqnarray}

There is a boundary condition                                          %
$\theta_{g\rightarrow \pm \infty}=\pi/2$ if the chain is long
enough, that means if $KN>>L$, with $N$ being the number of magnetic 
ions of the chain. Further, the above symmetry conditions imply 
that only the sites with $g\ge 0$ must be considered.

For the  double exchange case, the same set of equations is
obtained but the term in $L$ in equation (\ref{ eq:2}) must be changed by
the standard effective hopping of the double exchange model,
$-t\cos\left(\frac{\theta_{-1}+\theta_{0}}{2}\right)$ 
\footnote{Note that this is a two site problem, and therefore no 
additional phases are needed.}, 
and the constant
term $-t$ must be changed by $-2L$. Using the symmetry
condition $\theta_{-1}=\theta_{0}$, this is equivalent to change $L$ by
$t/2$ in equation (\ref{ eq:3}). Therefore both cases can be treated on
the same footing. 

Following the paper by N\'eel \cite{Nee67}, we look for a differential %
equation for the d-spin distortion. If we were dealing with a
ferromagnetic d-d exchange, the set of equations (\ref{ eq:3}) would
describe a domain wall. To obtain a differential equation for our
problem an additional step is needed. Instead of working with the 
angles $\theta_{g}$, we perform a rotation of an angle $\pi$ around 
each $x$-axis if the site is an odd site, and no rotation if the site
is an even site.  This corresponds to make the following changes in
the angles: $\theta_{2n}\rightarrow\theta_{2n},\quad
\theta_{2n+1}\rightarrow\pi-\theta_{2n+1}$.
We assume that the length of variation of the angle $\theta$ is larger
than the lattice constant. We treat $\theta$ as a continuous function
over the $y$-axis and perform a power expansion in the lattice constant. 
Then a differential equation is obtained. Taking into account the above 
symmetries, we have to solve only for the positive semi-axis. Further, 
we divide the problem in two parts. For $y>0$, we have:

\begin{equation}\label{ eq:6}
J\frac{d^{2} \theta}{dy^{2}}+ K \sin 2\theta =0
\end{equation}
For $y=0$ and using again the symmetries, we obtain:

\begin{equation}\label{ eq:7}
J\left.\frac{d \theta}{dy}\right|_{y=0}+(J+K) \sin 2\theta_{0}
-L\sin\theta_{0}=0
\end{equation}
where $\theta_{0}=\theta\left(y=0\right)$. This is the sine-Gordon 
equation with a boundary condition at the origin. 

We solve equation (\ref{ eq:6}) with the boundary condition that  
$\theta_{y\rightarrow +\infty}\rightarrow\pi/2$. Multiplying
by $2\frac{d \theta}{dy}$ and integrating once, we obtain:

\begin{equation}\label{ eq:8}
J \left(\frac{d \theta}{dy}\right)^{2}-K \cos 2\theta=C
\end{equation}
where $C$ is the integration constant. Following reference 
\cite{Aha93}, we made:

\begin{eqnarray}\label{ eq:9}
C&=&J \left.\left(\frac{d \theta}{dy}\right)^{2}\right|_{y=0}
-K \cos 2\theta_{0}\nonumber\\
&=&\frac{1}{J}
\left[L\sin\theta_{0}-\left(J+K\right)\sin2\theta_{0}\right]^{2}
-K\cos2\theta_{0}
\end{eqnarray}
The last step is to take into account equation (\ref{ eq:7}).
Now we made the following change:

\begin{equation}\label{ eq:11}
\sin \theta=\frac{f\left(y\right)-1}{f\left(y\right)+1}
\end{equation}
The differential equation (\ref{ eq:8}) is cast into the form:

\begin{eqnarray}\label{ eq:12}
\frac{1}{\left(f+1\right)^{2}}
\left[\frac{1}{f}\left(\frac{df}{dy}\right)^{2}+\right.
\left.\frac{2K}{J}\left(f-1\right)^{2}\right]=\nonumber\\
\left[\frac{L}{J}\frac{f_{0}-1}{f_{0}+1}\right.
\left.-4\frac{J+K}{J}\frac{f_{0}-1}{\left(f_{0}+1\right)^{2}}\sqrt{f_{0}}\right]^{2}
\nonumber\\
+\frac{2K}{J}\left(\frac{f_{0}-1}{f_{0}+1}\right)^{2}
\end{eqnarray}
where we define $f_{0}=f\left(y=0\right)$. The solution of equation    %
(\ref{ eq:12}) is:

\begin{equation}\label{ eq:13}
f\left(y\right)=\exp\left(a+b y\right)
\end{equation}
with $a,\,b$ being two real constants to be determined. Note that      %
$a,\,b>0$ guarantees that the angle $\theta$ lies in the range
$(0,\pi/2)$ for $y$ belonging to the domain $[0,+\infty)$. Also 
note that at the infinite $\theta$ goes to $\pi/2$, as required.
The differential equation is identically satisfied if:

\begin{equation}\label{ eq:14}
b^{2}=\frac{8K}{J}
\end{equation}
and:

\begin{equation}\label{ eq:15}
\sqrt{\frac{8K}{J}}\frac{\sqrt{f_{0}}}{f_{0}-1}=\frac{L}{J}
-4\frac{J+K}{J}\frac{\sqrt{f_{0}}}{f_{0}+1}
\end{equation}
which is numerically solved to obtain $f_{0}$, or alternatively $a$.   % 
In reference \cite{Aha93}, the solution of the equation (\ref{ eq:6})
with the boundary condition (\ref{ eq:7}) is treated in detail.   

\begin{figure}
\begin{center}
\epsfig{file=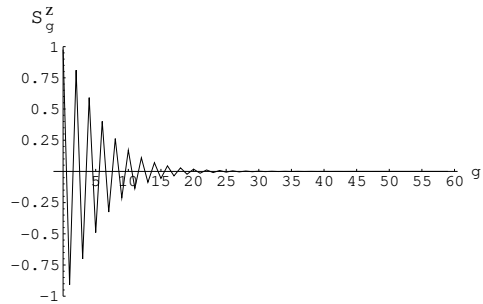,height=4 cm}
\end{center}
\caption{\label{ fig:1} Magnetization along $z$-axis, $S^{z}_{g}$, 
(in units of $S$) with $L=3$ and anisotropy $K=2.5\cdot10^{-2}$
(both in $J$ units). Analytical calculation.} 
\end{figure}

In figure~\ref{ fig:1} we plot the magnetization along the $z$-axis,
$S^{z}_{g}$, for the  values of $L=3,\,K=2.5\cdot10^{-2}$ (both in $J$
units). This is obtained by inverting the above changes to the original
angles. 

We calculate the total magnetization that appears in the d-spins system
along the $z$-axis. The result is show in table \ref{ tab:1}. As can be
seen, the existence of the magnetic distortion outside the localization
region of the charge carrier leads to a partial compensation in the 
total magnetization.

With the analytical solution one can easily calculate the radius of    %
the magnetic distortion. As can be seen from the solution, equations 
(\ref{ eq:11}, \ref{ eq:12}), the radius is not properly defined, 
because the distortion created by the conduction electron in the 
magnetic system of d-spins reaches the complete chain.
We choose to define the radius as the distance at which the straight
line with slope equal to $\left.\frac{d \theta}{dy}\right|_{y=0}$ and
passing throughout the origin at $\theta_{0}$, reaches the value
$\pi/2$. This underestimates the radius, but has the obvious advantage
that in this way, the radius depends only on the value of the constant
$a$, or alternatively on the angle $\theta_{0}$:

\begin{equation}\label{ eq:16}
R=\left(\frac{\pi}{2}-\theta_{0}\right)\sqrt{\frac{J}{2K}}
\sec\theta_{0}
\end{equation}
Also note that the value of $\theta_{0}\approx0$ and therefore
$R\approx\pi\sqrt{J/8K}$. We obtain $R=6.18981$ (in lattice 
constant units) for $K=2.5\cdot10^{-2}$ (in $J$ units).

With the analytical solution it is also easy to calculate the energy of   %
the ferron. The equation for the energy, equation (\ref{ eq:2}),
contains also the energy for the system of d-spins even in the case
with no conduction electron present. This energy has a value of
$E_{0}=-(J+K)N$. We define the energy of the magnetic polaron as 
$E-E_{0}$. As the transformation that we perform to obtain the
differential equation (\ref{ eq:8}) is a canonical transformation, we
can use it again to calculate the energy. Taking this into account we 
transform the equation (\ref{ eq:2}) and make again a power expansion
in the lattice constant:

\begin{equation}\label{ eq:18}
E=E_{\text{core}}-2\int^{\frac{N}{2}}_{0}dy
\left\{J\left[1-\frac{1}{2}\left(\frac{d\theta}{dy}\right)^{2}\right]+K\sin^{2}\theta\right\}
\end{equation}
where 
$E_{\text{core}}=J\left(1+\cos2\theta_{0}\right)                       %
-2L\cos\theta_{0}-t$
\footnote{As $t$ does not enter in the equation to determine the magnetic
structure (\ref{ eq:3}), its numerical value is set to zero. Here it is
shown only for completeness. Also for the double exchange case, one has 
to change $L$ by $t/2$, and conversely.}.   
Therefore the polaron energy is:

\begin{equation}\label{ eq:21}
E_{\text{pol}}=E-E_{0}=E_{\text{core}}
+\sqrt{8JK}\left(1-\sin\theta_{0}\right)
\end{equation}
as expected from the result for the ferromagnetic domain wall. Also
note again that the value of $\theta_{0}$ is small and therefore
$E_{\text{pol}}\approx-2\left(L-J\right)-t+\sqrt{8JK}$, not only in 
the case of wide-band, but also in the case of double exchange 
semiconductor.

The motivation for the energy calculation is to demonstrate that the   %
ground state energy of a bound ferron with a magnetization
compensating region is much lower than the ground state energy of a
bound ferron without such a long range distortion. In table \ref{ tab:1}, we 
present the result for the ground state energy 
of the magnetic polaron calculated from the equation (\ref{ eq:21}).
We also show for comparison the result in the case without 
compensating region. This latter was calculated solving the
set of equations (\ref{ eq:3}) numerically, and imposing 
$\theta_{1}=\pi/2$. As expected the energy coming from equation
(\ref{ eq:21}) is lower, meaning that a true bound ferron is much
more stable than bound ferrons considered previously in the literature,
as it was once again anticipated by Nagaev in \cite{Nag01b}.

%\begin{center}
\begin{table}
\begin{tabular}{|r|c|c|c|} 
\hline
& Ground state energy & $\theta_{0}$ & $M_{z}$\\
\hline\hline
With comp. reg. & -3.60101 & 0.22011 & 1.00019$S$ \\
\hline
Without comp. reg. & -2.68811 & 0.63688 & 1.60791$S$ \\ 
\hline
\end{tabular}
\caption{\label{ tab:1} Ground state energies (in $J$ units), canting 
angle in the core, $\theta_{0}$, and total magnetization along
$z$-axis (in $S$ units), $M_{z}$, for a 1D magnetic polaron with,
and without compensating region, using $L=3,\,K=2.5\cdot 10^{-2}$
(both in $J$ units).}
\end{table}
%\end{center}

As can be seen from the equation (\ref{ eq:21}), the main part of the  %
energy of the magnetic polaron is concentrated in the core, that means
at the sites $g=-1,0$, where the charge carrier is trapped. The energy 
in the compensating region, that means outside the sites $g=-1,0$, is 
very small and positive. This seem to be in contradiction with the 
previous discussion, in which the presence of a compensating region 
was presented as energetically favored. The physical explanation of 
this behavior is the following. The main energy scale in the problem 
that is coupled to the magnetic ordering is $L$, the s-d interaction, 
in a wide-band  semiconductor. It tends to put the d-spins in the
core as parallel to the $z$-axis as possible. But the d-spins in the
core are connected to the rest of the chain through the d-d exchange 
term. The role of the compensating region is to isolate the d-spins 
in the core from the rest of the chain. This allows the d-spins in the 
core to be parallel to the $z$-axis with a high gain in s-d exchange 
energy ($E_{\text{core}}$ is high and negative) and a little loss due 
to the perturbation of the antiferromagnetic ordering on the remaining 
part of the chain (the energy of the compensating region is positive 
but small). To better explain this point, we also show in table 
\ref{ tab:1}, the canting angle of the d-spins of the core, $\theta_{0}$, 
for a bound ferron with compensation region, and a bound ferron without
compensating region. As can be seen, the presence of the magnetization
compensation region with the structure described above strongly reduces
the value of canting angle. The role of the anisotropy is simply to
set a scale for the radius of the magnetic distortion in the d-spin 
system. 

Apart for the solution treated here, there is another possible solution
for the d-spin structure
\footnote{K. I. Kugel, and A. O. Sboychakov (private communication)}.
This second solution is characterized by the absence of extended spin
distortions. It corresponds to rotate by $\pi$ the d-spins of one half,
say the negative one, $g<0$, of the chain. This make that the
magnetization in the trapping region will be directed along $x$-axis,
instead of $z$-axis. Mathematically, it corresponds to choose a different
symmetry condition for the d-spin system, namely $S^{y}_{g}=0$, and 
$S^{x}_{g}=S^{x}_{-\left(g+1\right)}$. This solution saves more anisotropy
energy and allows for a lower canting angle in the core, being at first
instance the ground state in the range of parameters chosen here. However, as 
the Heisenberg model has to be solved over a compactified ring, that is, the 
d-spin at $g\longrightarrow +\infty$ is linked to the d-spin at 
$g\longrightarrow -\infty$, one has to add an extra energy $2J$ to the polaron
energy, equation (\ref{ eq:18}), resulting that for the case of an isolated 
ferron the ground state energy always corresponds to the solution treated 
here. In the case of a finite density of ferrons, this condition at infinity 
can be skipped and both solutions would be possible. The finite density case 
is left for further work.

The three-dimensional case cannot be solved analytically. We only      %
give a simple estimate for the radius of the distortion for an 
antiferromagnetic coupling. We use the three-dimensional analogue of 
the Hamiltonian (\ref{ eq:1}). The case $K'=0$ is treated in reference 
\cite{DeG60}. To introduce the anisotropy we use a simple variational 
method. We assume that the distortion angle is given by the 
spherically-symmetrical asymptotic solution of De Gennes,
$\theta\left(r\right)\sim1/r^2$ for $r<R$, and that there is no
distortion for $r>R$. The optimal value of $R$ is obtained by 
minimization. For $R$ large, we obtain
$R\sim\left(J/K\right)^{\frac{1}{6}}$. This provides an order of 
magnitude for the radius of the distortion. 
With this estimate for the radius of the distortion is now clear that,
in real three-dimensional materials and over experimental doping ranges,
the (one-electron) ferrons treated here must overlap, forming magnetic
droplets with a number of conduction electrons larger than one, as
reported in the experiment. 
Therefore it is expected 
that the main conclusions of this article hold for the 
three-dimensional case, although further calculations are needed.

To summarize, we have found the detailed structure of the
one-dimensional d-spin system in the region surrounding of a bound
ferron, completing the previous results of \cite{Nag01b,DeG60,Nag67}.
The main result is the appearance of a new length scale, namely the 
extent of the magnetic distortion created by the charge carrier. The 
existence of this distortion makes the ferron more stable. This may
determine, together with the Coulomb interaction, the spatial 
distribution of magnetic droplets and their coupling in
antiferromagnetic semiconductors, such as underdoped manganites. Also
it is related to the onset of the electronic phase separation at
the very low doping range observed in these compounds.

\begin{acknowledgments}
The authors are deeply indebted to Prof. D. I. Khomskii for proposing
this problem and his encouragement and help during its realization. We
also acknowledge to Profs. E. Dagotto, K. I. Kugel, and A. O. Sboychakov 
for useful discussions.
\end{acknowledgments}

\bibliography{article3}

\end{document}